
\input phyzzx
\singlespace
\twelvepoint
\REF\bv{N. Berkovits and C. Vafa, {\sl On the Uniqueness of String Theory},
                 preprint HUTP-93/A031
                 , KCL-TH-93-13, hep-th/9310170 (1993).}
\REF\oth{J.M. Figueroa-O'Farrill, {\sl On the Universal String Theory},
                  preprint QMW-PH-93-29, hep-th/9310200 (1993);
                  {\sl The Mechanism behind the Embeddings
                  of String Theories},
                  preprint QMW-PH-93-30, hep-th/9312033;\hfill
\break       H. Ishikawa and M. Kato,

                  {\sl Note on N=0 string as N=1 string},
                   preprint UT-Komaba/93-23, hep-th/9311139;\hfill
\break       N. Ohta and J.L. Petersen, {\sl N=1 from N=2 Superstring},
                   preprint NBI-HE-93-76, hep-th/9312187.}
\REF\BOP{F. Bastianelli, N. Ohta and J.L. Petersen,
            {\sl Toward the Universal Theory
		 of Strings}, preprint NORDITA-94/6 P, hep-th/9402042.}
\REF\BFW{N. Berkovits, M. Freeman and P. West,
                  {\sl A W-String Realization of the Bosonic String},
                  preprint KCL-TH-93-15, hep-th/9312013.}
\REF\kli{K. Li, {\sl Linear $w_N$\ Gravity}, preprint CALT-68-1724 (1991).}
\REF\witten{E. Witten, {\sl Surprises with Topological Field Theories},
in the proceedings of {\it Strings '90} ed. by R. Arnowitt {\it et al.}

                       (World Scientific) (1991).}
\REF\hoso{S. Hosono, {\sl Int. J. Mod. Phys.} {\bf A7} (1992) 5193.}
\REF\Ma{E. Martinec, {\sl Nucl. Phys.} {\bf B281} (1987) 157.}
\REF\ik{H. Ishikawa and M. Kato, {\sl Note on N=0 string as N=1 string},
                   preprint UT-Komaba/93-23, hep-th/9311139.}
\REF\fms{D. Friedan,E. Martinec and S. Schenker,

                           {\sl Nucl. Phys.} {\bf B271} (1986) 93.}
\pubnum{OU-HET 187}
\titlepage
\title{A Universal $w$\ String Theory}
\author{Hiroshi KUNITOMO\foot{
e-mail address: kunitomo@osklns.kek.ac.jp}\llap,
Makoto SAKAGUCHI\foot{
e-mail address: gu@oskth.kek.ac.jp}
\break and
\break Akira TOKURA\foot{
e-mail address: atoku@oskth.kek.ac.jp}}
\address{Department of Physics, Osaka University
\break Toyonaka, Osaka 560, JAPAN}
\abstract{
It has been shown that there is a sequential embedding structure
in a $w_N$\ string theory based on a linearized $W_N$\ algebra.
The $w_N$\ string theory is obtained as a special realization of
the $w_{N+1}$\ string.
The $w_{\infty}$\ string theory is a universal string theory in this sense.
We have also shown that there is a similar sequence for $N=1$\

string theory.
The $N=1\ w_N$\ string can be given as a special case
of the $N=1\ w_{N+1}$\ string.
In addition, we show that the $w_3$\ string theory
is obtained as a special realization of
the $N=1\ w_3$\ string. We conjecture that the $w_N$\ string
can be obtained as a special $N=1\ w_N$\ string for general $w_N$.
If this is the case, $N=1\ w_{\infty}$\ string theory is more universal
since it includes both $N=0$\ and $N=1$\ $w_N$\ string theories.
}
\endpage
\chapter{Introduction}

The symmetry breaking is one of the important concept in physics.
On the unified theory view point, it can be used to lead variety of
theories from a unified theory which has
higher symmetry. This idea has another advantage since the higher symmetry
constrains
the theory more strongly. If there exists a unique unified theory,
it should have a
maximum symmetry. It is a fascinating idea that there is such a
{\it universal theory}
from which all the possible theories can be lead.

In the string theory, it has been recently found that this kind of
unification is
possible\rlap.\refmark{\bv,\oth} $N=0$\ ($N=1$) string theory can be lead
from
a special $N=1$\ ($N=2$) string theory. This is such a new kind of symmetry
breaking that the physical subspace has less symmetry than the whole
Hilbert space
since some members of a multiplet are decoupled from the physical space as
a BRST quartet.
This result has been able to be extended to $N=4$\ string\refmark{\BOP}
which is
a maximal number of (linear) supersymmetries in two dimensions. This theory
is
a candidate of the universal theory which leads all the superstring
theories.

There is another sequence of theories that is one of $W_N$\ strings.
While embedding of bosonic ($W_2$) string in the $W_3$\ string has been
found in
Ref. [\BFW], it is complicated due to the nonlinear nature of the
$W_3$\ algebra and
is difficult to extend to general $W_N$\ theories.
In this paper, we consider linearized $w_N$\ algebra, called linear
$w_N$\ algebra,
and corresponding string theory which has been obtained in Ref. [\kli].
This has an advantage that it is easy to investigate general structure of
embeddings.
The theory which has a maximum symmetry is the $w_{\infty}$\ string theory
and it is a universal theory of this sequence.
Unfortunately the sequential embeddings of this case are not general.
The $w_N$\ string theory included in the $w_{N+1}$\ string has the special
form
which is used to realize further symmetry breaking to the $w_{N-1}$\ string.
Only the bosonic ($w_2$) string theory is embedded in the general form.

We have two sequences of embeddings
labeled by the number of supersymmetries and
by the number of higher dimensional generators.
The $w_{\infty}$\ string theory is not universal in a sense that it does
not include
the sequence of superstrings. We further consider the
$N=1\ w_{N+1}$\ string theory
and show that this includes the $N=1\ w_N$\ string.
It is also shown that there is another embedding structure
related these two sequences.
The $w_3$\ string can be obtained by a special $N=1\ w_3$\ string.
If this embedding structure holds for general $w_N$\ string,
$N=1\ w_{\infty}$\ string theory is more universal which include a part of
the sequence of supersymmetries.

The linearization of $W_N$\ algebra is not unique and there is
another linearized $w_N$\ algebra\refmark{\hoso} which we call the abelian
$w_N$\ algebra. This linearization is such a simple that
all the higher dimensional generators commute.
We can repeat all the investigations for another $w_N$\ string theory
based on the abelian $w_N$\ algebra.
In this case, the general $w_N$\ string is obtained as a special $w_{N+1}$\
string, which is different to the linear $w_N$\ string.
However, the geometrical meaning of the universal theory is not clear
since the $N\rightarrow\infty$\ limit of the abelian $w_N$\ algebra is not
the area preserving diffeomorphisms.

This paper is organized as follows.

We explain the linear $w_N$\ algebra in  \S 2.
The string theory is constructed based on the  linear $w_N$\ algebra.
The BRST operator is constructed for the $w_N$\ string theory.
For discussing the physical vertex operators,
we define the invariant vacuum under the $sl(N,{\bf R})_c$\
which is the finite subalgebra of the linear $w_N$\ algebra.

In \S 3, we show that the $w_N$\ string theory can be obtained by a special
realization of the $w_{N+1}$\ string. By a similarity transformation,
the BRST operator of $w_{N+1}$\
string is transformed into the sum of the BRST operators of
the $w_N$\ string and a topological theory.
Since the cohomology of the topological theory is trivial,
the cohomology of the $w_{N+1}$\ string
is the same with the one of the $w_N$\ string.
We also prove that amplitudes of $w_{N+1}$\ string coincide
with those of $w_N$\ string.

The $N=1\ w_N$\ string theory is introduced in \S 4,
where the $N=1\ w_N$\ algebra and the BRST operator are defined.
We show, in \S 5, that there is a similar sequence to $N=0$\
for $N=1\ w_N$\ string theory. The $N=1\ w_N$\ string can be obtained by a
special
realization of $N=1\ w_{N+1}$\ string.

In the above argument of the linear $w_N$\ string theory,
the $w_N$\ string obtained by the special $w_{N+1}$\ string is not the
general one.
In \S 6, it is shown that
we can obtain the general $w_N$\ string theory if we take another
linearization
of the $W_N$\ algebra, the abelian $w_N$\ algebra.
We repeat the discussions in \S 2-5 for the abelian $w_N$\ algebra.
The general $w_N$\ string is obtained as a special $w_{N+1}$\ string
while the algebra obtained by
$N\rightarrow \infty$\ limit is different from the area preserving
diffeomorphisms.

In \S 7, we discuss another embedding of the string:
the $w_N$\ string can be obtained as a special $N=1\ w_N$\ string.
The explicit construction of embedding is given for the case of the
$w_3$\ string.
If this embedding structure exists for general $w_N$,
the $N=1\ w_{\infty}$\ string is more universal
theory than $w_{\infty}$\ string.

\S 8 is devoted to the discussions.

\chapter{$w_N$\ string theory}

In an analogous way to Ref. [\bv], it has been found that the bosonic
($W_2$) string
theory can be obtained from the $W_3$\ string theory by taking a special
realization
of $W_3$\ algebra\rlap.\refmark{\BFW} It is natural to generalize this to
string
theories with higher symmetry, $W_N$\ string theories. The general
$W_N$\ algebra is,
however, too complicated to construct such a special realization.
This complexity come
from the fact that the $W_N$\ algebra is nonlinear. In order to investigate
the sequence of
embeddings in $W_N$ string theories, it is better to use simpler algebra
which can be
constructed generally.
In this section, we consider a linearized analog of the
$W_N$\ algebra which we call the linear $w_N$\ algebra.
It is explained that the linear $w_N$\ algebra and
the string theory constructed based on it.

The $w_N$\ algebra is generated by currents
$w_i(z)\ (i=0,\cdots ,N-2)$\ which have dimension $i+2$.
The first current of them
is the stress tensor $w_0(z)=T(z)$.
A natural linearization of the  $W_N$\ algebra was obtained
in Ref. [\kli] as
$$
\eqalignno{
w_i(z)w_j(w)&\sim {{c\over2}\delta_{i,j}\delta_{i,0}\over (z-w)^4}+
{(i+j+2)w_{i+j}(w)\over (z-w)^2}+{(i+1)\partial w_{i+j}(w)\over z-w},\cr
&&{\rm for } \ i+j\le N-2,\cr
w_i(z)w_j(w)&\sim 0,&{\rm for }\  i+j>N-2.\cr
&&\eqname\linearw\cr
}
$$
This algebra becomes $w_{\infty}$, area preserving diffeomorphisms,
by taking the limit $N\rightarrow \infty$.

The $w_N$\ string theory is constructed by using the linear $w_N$\ algebra
and fermionic ghosts $(b_i^{gh}(z),\ c_i^{gh}(z)),\ i=0,\cdots ,N-2$.
These ghost fields have dimensions $(i+2,-i-1)$\ and satisfy the OPE
$$
\eqalignno{
c_i^{gh}(z)b_j^{gh}(w)&\sim {\delta_{i,j}\over z-w}.
&\eqname\bc\cr
}
$$
The Hilbert space is direct product of the representation space of the
linear $w_N$\
algebra and the ghost Fock space. The physical subspace is defined
by BRST operator
$$
\eqalignno{
Q_{BRST}^{(N)}&=\oint{dz\over 2\pi i}\sum^{N-2}_{i=0}c_i^{gh}
\big(w_i+{1\over 2}w_i^{gh}\big),
&\eqname\BRST\cr
}
$$
where the ghost's generators $w_i^{gh}(z)$\ are defined by
$$
\eqalignno{
w_i^{gh}(z)&=\sum^{N-i-2}_{j=0}
\big(-(i+j+2)b_{i+j}^{gh}\partial c_j^{gh}
-(j+1)\partial b_{i+j}^{gh}c_j^{gh}\big).
&\eqname\linw\cr
}
$$
This BRST operator is nilpotent if the matter system
has the central charge
$$
\eqalignno{
c(N)&=\sum^N_{j=2}2(6j^2-6j+1)\cr
&=2(N-1)(2N^2+2N+1).
&\eqname\cent\cr
}
$$

For discussing the physical operators,
let us introduce mode expansions as follows.
$$
\eqalignno{
w_i(z)&=\sum_n w_{i,n}z^{-n-i-2},\cr
b_i^{gh}(z)&=\sum_n b_{i,n}^{gh}z^{-n-i-2},\cr
c_i^{gh}(z)&=\sum_n c_{i,n}^{gh}z^{-n+i+1}.
&\eqname\mode\cr
}
$$
By means of these mode expansion, the linear $w_N$\ algebra are written as
$$
\eqalignno{
[w_{i,n},w_{j,m}]&=\{(j+1)n-(i+1)m\}w_{i+j,n+m}+
{c\over 12}n(n^2-1)\delta_{i,j}\delta_{i,0}\delta_{n+m,0},\cr
&&{\rm for}\ i+j \le N-2,\cr
[w_{i,n},w_{j,m}]&=0,& {\rm for}\ i+j > N-2.\cr
&&\eqname\lenwmod\cr
}
$$

The linear $w_N$\ algebra has finite subalgebra which is generated
by operators
$$
\eqalignno{
&\oplus^{N-2}_{i=0}\{\oplus^{i+1}_{n=-i-1}w_{i,n}\}.
&\eqname\fsub\cr
}
$$
We call this algebra $sl(N,{\bf R})_c$\ since it is obtained by
a group contractions of $sl(N,{\bf R})$\rlap.\refmark{\kli}
The $sl(N,{\bf R})_c$\ invariant vacuum is defined by
$$
\eqalignno{
&\ket{0}=\ket{0}_{mat}\otimes\ket{0}_{gh},\cr
&w_{i,n}\ket{0}_{mat}=0,\qquad {\rm for}\ n\ge -i-1,\cr
&b_{i,n}^{gh}\ket{0}_{gh}=0,\qquad {\rm for}\ n\ge -i-1,\cr
&c_{i,n}^{gh}\ket{0}_{gh}=0,\qquad {\rm for}\ n > i+1,\cr
&\qquad\qquad\qquad\qquad{\rm for}\ i=0,\cdots,N-2.
&\eqname\slvac\cr
}
$$
The physical state is obtained, up to picture changing explained later, as
$$
\eqalignno{
\ket{phys}&=V_{mat}(0)\ket{0}_{mat}\otimes
\prod^{N-2}_{i=0}\prod^{i+1}_{n=1}c_{i,n}^{gh}\ket{0}_{gh},
&\eqname\physs\cr
}
$$
or equivalently, the physical vertex operator has the form
$$
\eqalignno{
V_{phys}(z)
&=\prod^{N-2}_{i=0}\prod^{i}_{n=0}\partial^nc_i^{gh}(z)V_{mat}(z).
&\eqname\physv
}
$$
Here $V_{mat}(z)$\ is a primary field of the matter $w_N$\ algebra
which must have the dimensions ${N\over 6}(N^2-1)$.

Before closing this section, we should comment on the number of the
$w_N$\ moduli.
The moduli space of $w_N$\ algebra can be defined by the moduli space
of flat
$sl(N,{\bf R})_c$\ connections\refmark{\witten,\kli,\hoso}
$$
{\cal M}_{g,n}={\rm Hom}(\pi_1(\Sigma_{g,n}) , SL(N,{\bf R})_c)/\sim
$$
for the genus $g$\ and $n$\ punctured surface.
The dimension of this moduli space is obtained by an index theorem as
$$
\eqalignno{
dim{\cal M}_{g,n}=&(N^2-1)(g-1)+N(N-1)n/2 \cr
=&\ 3(g-1)+n  \cr
&+5(g-1)+2n  \cr
&+\cdots  \cr
&+(2N-1)(g-1)+(N-1)n,
&\eqname\moduli\cr
}
$$
where the number on the $i$-th line in the last equality is the number of
the moduli
comes from the spin $i+1$\ gauge field\rlap.\refmark{\Ma}

\chapter{$w_N\subset  w_{N+1}$}

In the previous section, we have constructed the $w_N$\ string theory
based on
the linear $w_N$\ algebra. Let us show, in this section,
that this linear $w_N$\ string theory is interpreted as the special
case of the linear $w_{N+1}$\ string theory.

Consider the general critical bosonic ($w_2$) string theory represented by
the stress tensor $T(z)$\ with $c=26$.
By adding bosonic {\it matter} fields
 $(\beta_i(z),\gamma_i(z)),\ i=1,\cdots,N-1$,
we can construct generators of the $w_{N+1}$\  algebra
$\{w_i(z)\},\ i=0,\cdots,N-1$\ as
$$
\eqalignno{
w_i(z)&=\delta_{i,0}T+i\beta_i
+\sum^{N-i-1}_{j=1}
\big(-(i+j+2)\beta_{i+j}\partial\gamma_j
-(j+1)\partial\beta_{i+j}\gamma_j\big).
&\eqname\lsw\cr
}
$$
Here the additional fields have dimensions $(i+2,-i-1)$\ and
satisfy the OPE
$$
\eqalignno{
\gamma_i(z)\beta_j(w)&\sim {\delta_{i,j}\over z-w}.
&\eqname\opeb\cr
}
$$
This realization can be interpreted as being constructed by
the $w_N$\ string
defined by generators $\{\tilde w_i\},\ i=0,\cdots,N-2$\ and
$(\beta_{N-1}(z),\gamma_{N-1}(z))$,
where $\tilde w_i$\ is obtained by removing
$(\beta_{N-1}(z),\gamma_{N-1}(z))$\
in \lsw. However, this $w_N$\ string is the special one realized by means
of $T(z)$\ and $(\beta_i(z),\gamma_i(z)),\ i=1,\cdots,N-2$. Only the
$w_2$\ (bosonic)
string theory is included as a whole. By this construction,
therefore, we can only show that
the linear $w_{N+1}$\ string is including special realizations of
the linear $w_i$\
string $(i=3,\cdots,N)$\ and the general $w_2$\ string.

Since the central charge of this matter system is equal to the critical
value $c(N+1)$,
the $w_{N+1}$\ string theory is obtained by adding the fermionic ghosts
$(b_i^{gh}(z),c_i^{gh}(z))\ i=1,\cdots ,N-1$\ as explained in \S 2.
We can prove the coincidence of the cohomology of this special realization
of $w_{N+1}$\ string and of the $w_N$\ string by giving a similarity
transformation\refmark{\ik} defined by
$$
\eqalignno{
R&={1\over N-1}\oint {dz\over 2\pi i}\sum^{N-2}_{i=0}
c_i^{gh}\Big[(N+1)b_{N-1}^{gh}\partial\gamma_{N-i-1}+
(N-i)\partial b_{N-1}^{gh}\gamma_{N-i-1}\Big].
&\eqname\simtwo\cr
}
$$
The BRST operator of the $w_{N+1}$\ string $Q_{BRST}^{(N+1)}$\ is
transformed by this similarity
 transformation into the sum of the BRST operators of
the $w_N$\ string $Q_{BRST}^{(N)}$\ and a topological theory $Q_{top}$.
$$
\eqalignno{
e^RQ_{BRST}^{(N+1)}e^{-R}&=Q_{BRST}^{(N)}+Q_{top},
&\eqname\BRSTone\cr
}
$$
where
$$
\eqalignno{
Q_{top}&=\oint {dz\over 2\pi i}(N-1)c_{N-1}^{gh}\beta_{N-1}.
&\eqname\topone\cr
}
$$
Therefore the physical states of the $w_{N+1}$\ string are given by
the tensor product of the physical states of the $w_N$\ string and
the topological theory. The further discussions hereafter are given
in the transformed theory.
The results can be inversely transformed if one want to get those in the
original form.

The cohomology of the topological theory
is trivial since the topological BRST operator has the bilinear
form and thus all the fields $(\beta_{N-1}(z),\gamma_{N-1}(z))$\ and
 $(b_{N-1}^{gh}(z),c_{N-1}^{gh}(z))$\ make the quartet and are decoupled
from the physical
subspace. Only the physical state in the topological sector is the
physical vacuum
defined later. To explain this further,
we note that the $sl(N+1,{\bf R})_c$\ vacuum split into the direct product
of the $sl(N,{\bf R})_c$\ vacuum in the $w_N$\ sector and the vacuum in the
topological
sector. The former is defined in eq. \slvac\  and the latter is obtained by
$$
\eqalignno{
\beta_{N-1,n}\ket{0}_{top}&=0,\qquad{\rm for}\ n\ge -N, \cr
\gamma_{N-1,n}\ket{0}_{top}&=0,\qquad{\rm for}\ n> N, \cr
b^{gh}_{N-1,n}\ket{0}_{top}&=0,\qquad{\rm for}\ n\ge -N, \cr
c^{gh}_{N-1,n}\ket{0}_{top}&=0,\qquad{\rm for}\ n> N,
&\eqname\topvac\cr
}
$$
where we introduce the mode expansion of
$(\beta_{N-1}(z),\gamma_{N-1}(z))$\ as
$$
\eqalignno{
\beta_{N-1}(z)&=\sum_n \beta_{N-1,n} z^{-n-N-1} \cr
\gamma_{N-1}(z)&=\sum_n \gamma_{N-1,n} z^{-n+N}.
&\eqname\modebg\cr
}
$$
In order to define the physical vacuum,
we must bosonize {\it the matter}\ fields
$(\beta_{N-1}(z),\gamma_{N-1}(z))$\ in a similar way to the bosonic ghost
fields
in the superstring theory\refmark{\fms}\foot{
Here we invert the role of
$(\eta_{N-1}(z),\partial\xi_{N-1}(z))$\ comparing with
the bosonic ghosts. This bosonization is convenient for this case as
becoming clear later.
}:
$$
\eqalignno{
\beta_{N-1}(z)&=e^{-\phi_{N-1}(z)}\eta_{N-1}(z),  \cr
\gamma_{N-1}(z)&=-\partial \xi_{N-1}(z)e^{\phi_{N-1}(z)}
&\eqname\boso\cr
}
$$
The mode expansions of these fields is defined by
$$
\eqalignno{
i\partial\phi_{N-1}(z)&=\sum_n \phi_{N-1,n} z^{-n-1},\cr
\eta_{N-1}(z)&=\sum_n \eta_{N-1,n} z^{-n-1},\cr
\xi_{N-1}(z)&=\sum_n \xi_{N-1,n} z^{-n}
&\eqname\modebo\cr
}
$$
The topological vacuum \topvac\ is defined by these bosonized fields as
$$
\eqalignno{
\phi_{N-1,n}\ket{0}_{top}&=0,\qquad {\rm for}\ n\ge 0,\cr
\eta_{N-1,n}\ket{0}_{top}&=0,\qquad {\rm for}\ n\ge 0,\cr
\xi_{N-1,n}\ket{0}_{top}&=0,\qquad {\rm for}\ n > 0.
&\eqname\slvacbo\cr
}
$$
The physical vacuum of the topological sector is defined by
$$
\eqalignno{
\ket{0}_{top}^{phys}&=\prod^{N-1}_{n=0}\partial^nc_{N-1}^{gh}(0)
e^{-N\phi_{N-1}(0)}\ket{0}_{top}.
&\eqname\physvac\cr
}
$$
Therefore the physical vertex operator of the $w_{N+1}$\ string is
obtained by
$$
\eqalignno{
V_{phys}^{(N+1)}(z)
&=(\prod^{N-1}_{n=0}\partial^nc_{N-1}^{gh}(z)e^{-N\phi_{N-1}(z)})
V_{phys}^{(N)}(z),
&\eqname\physv\cr
}
$$
where $V_{phys}^{(N)}(z)$\ is the physical vertex operator in the
$w_N$\ string.

In the remaining part of this section, we discuss that
the amplitudes of the $w_{N+1}$\ string coincide with those of the $w_N$\
string. This is nontrivial since the rules of calculating the amplitudes
are different for two theories.
For calculating the amplitudes,
we need to introduce the picture changing operator
$Z_{N-1}(z)$\ as in the superstring theory\rlap.\refmark{\fms}
It is obtained by
$$
\eqalignno{
Z_{N-1}(z)&=\{Q^{(N+1)}_{BRST},\xi_{N-1}(z) \}\cr
&=\{Q_{top},\xi_{N-1}(z) \}\cr
&=(N-1)c_{N-1}^{gh}(z)e^{-\phi_{N-1}(z)}.
&\eqname\pcop\cr
}
$$
The inverse picture changing operator,
which is needed to calculate the amplitudes,
is easily found as
$$
\eqalignno{
Z^{-1}_{N-1}(z)&={1\over{N-1}}e^{\phi_{N-1}(z)}b_{N-1}^{gh}(z),
&\eqname\invpcop\cr
}
$$
which satisfies
$$
\eqalignno{
\lim_{z\rightarrow w}Z^{-1}_{N-1}(z)Z_{N-1}(w)&=1,\cr
[Q_{BRST}^{(N+1)},Z^{-1}_{N-1}(z)]&=0.
&\eqname\invpct\cr
}
$$
If we assume that the rules of calculating the amplitudes of the
$w_N$\ string are given,
(the holomorphic part of)
the amplitudes of the $w_{N+1}$\ string can be defined by using the inverse
picture changing operator as
$$
\eqalignno{
{\cal A}^{(N+1)}_{g,n}&=\int \prod^{3g-3+n}_{i=1}dm_i
\VEV{{\cal O}\prod^{(2N+1)(g-1)+Nn}_{i=1}Z^{-1}_{N-1}(z_i)
\prod^{n}_{j=1}V_{phys}^{(N+1)}(z_j)
}_{N+1},
&\eqname\ampone\cr
}
$$
where we denote the operators needed to insert for calculating the
amplitudes
of the $w_N$\ string as {\cal O} .
The number of the inserted $Z^{-1}_{N-1}$\
is the same as the number of the spin $N+1$\ moduli explained in \S 2.
We can interpret this insertion as the result of
the integration of these moduli.
One can easily see that this is factorized to the product of
the amplitudes of the $w_N$\ string and of the topological theory.
It coincides
the amplitudes of the $w_N$\ string
since the part of the topological theory are
one always up to sign comes from ordering of the fermionic operators.
$$
\eqalignno{
{\cal A}^{(N+1)}_{g,n}&=\int \prod^{3g-3+n}_{i=1}dm_i
\VEV{{\cal O}\prod^{n}_{j=1}V_{phys}^{(N)}(z_j)
}_N\times \cr
&\qquad\VEV{
\prod^{(2N+1)(g-1)+Nn}_{i=1}(e^{\phi_{N-1}(z_i)}b_{N-1}^{gh}(z_i))
\prod^{n}_{j=1}(\prod^{N-1}_{n=0}
\partial^nc_{N-1}^{gh}(z_j)e^{-N\phi_{N-1}(z_j)})
}_{top}\cr
&=\int \prod^{3g-3+n}_{i=1}dm_i
\VEV{{\cal O}\prod^{n}_{j=1}V_{phys}^{(N)}(z_j)
}_N\cr
&={\cal A}^{(N)}_{g,n},
&\eqname\amptwo\cr
}
$$
where it is noted that the similarity transformation \simtwo\
acts not only on the topological sector but also on the $w_N$\ sector.
One can see however that it does not affect the $w_N$\
amplitudes due to the (anomalous) $\beta_i$-$\gamma_i$\ number conservation.

Therefore it is completed the proof that
$w_N$\ string theory can be obtained by a special realization of
$w_{N+1}$\ string.
We have shown that there is a sequence of the $w_N$\ string theory.
$w_N$\ string theory can be interpreted as a special $w_{N+1}$\
string theory. We can consider the $w_{\infty}$\ string theory as a limit,
which has a maximum symmetry in this sequence and is a universal theory
in this sense.

\chapter{$N=1\ w_N$\ string theory}

We have shown that the $w_{\infty}$\
string theory is a universal theory which includes
the $w_N$\ string theories
as a special case.
The $w_{\infty}$\ string, however,
does not include the superstring theories.
Next we consider the $N=1\ w_N$\ string theory
and the sequential embeddings
including $N=1$\ superstring.

We define the $N=1\ w_N$\ string theory in this section.
$N=1$\ extension of the linear $w_N$\ algebra is generated by
bosonic generators $\{w_i(z)\}$\ and fermionic generators
$\{v_i(z)\},\ i=0,\cdots,N-2$.
Here we denote that $w_0(z)=T(z)$\ and $v_0(z)=G(z)$.
The dimensions of the bosonic (fermionic) generators are $i+2$\ ($i+3/2$).
The OPE relations of the $N=1$\ linear $w_N$\ algebra
is given by
$$
\eqalignno{
w_i(z)w_j(w)&\sim {{c\over 2}\delta_{i,j}\delta_{i,0}\over (z-w)^4}+
{(i+j+2)w_{i+j}(w)\over (z-w)^2}+{(i+1)\partial w_{i+j}(w)\over z-w},\cr
w_i(z)v_j(w)&\sim {(i+j+{3\over 2})v_{i+j}(w)\over (z-w)^2}+
{(i+1)\partial v_{i+j}(w)\over z-w},\cr
v_i(z)v_j(w)&\sim {{2c\over 3}\delta_{i,j}\delta_{i,0}\over (z-w)^3}+
{2w_{i+j}(w)\over z-w},\qquad {\rm for}\ \ i+j\le N-2,\cr
w_i(z)w_j(w)&\sim w_i(z)v_j(w)
\sim v_i(z)v_j(w)\sim 0,\qquad {\rm for}\ \ i+j>N-2.
&\eqname\Nw\cr
}
$$

For constructing the string theory,
we now need both fermionic and bosonic ghosts
$(b_i^{gh}(z),\ c_i^{gh}(z))$\ and $(\beta_i^{gh}(z),\ \gamma_i^{gh}(z)),\
i=0,\cdots N-2$. The dimensions of these ghost
fields are $(i+2,-i-1)$\ and $(i+3/2,-i-1/2)$.
The BRST operator is constructed as
$$
\eqalignno{
Q_{BRST}^{N=1(N)}&=\oint{dz\over 2\pi i}\sum^{N-2}_{i=0}\Big(c_i^{gh}
\big(w_i+{1\over 2}w_i^{gh}\big)\cr
&\qquad\qquad\qquad
 -\gamma_i^{gh}\big(v_i+{1\over 2}v_i^{gh}\big)\Big),
&\eqname\BRSTtwo\cr
}
$$
where
$$
\eqalignno{
w^{gh}_i(z)&=\sum^{N-i-2}_{j=0}
\Big(-(i+j+2)b_{i+j}^{gh}\partial c_j^{gh}
-(j+1)\partial b_{i+j}^{gh}c_j^{gh}\cr
&\qquad\qquad\qquad-(i+j+{3\over2})\beta_{i+j}^{gh}\partial\gamma_j^{gh}
-(j+{1\over2})\partial\beta_{i+j}^{gh}\gamma_j^{gh}\Big),\cr
v^{gh}_i(z)&=\sum^{N-i-2}_{j=0}
\Big(2b_{i+j}^{gh}\gamma_j^{gh}
-(i+j+{3\over2})\beta_{i+j}^{gh}\partial c_j^{gh}
-(j+1)\partial\beta_{i+j}^{gh}c_j^{gh}\Big).\cr
&
&\eqname\ghwv\cr
}
$$
The critical central charge is now
$$
\eqalignno{
c^{N=1}(N)&=\sum^N_{j=2}\{2(6j^2-6j+1)
-2(6(j-{1\over2})^2-6(j-{1\over2})+1)\}\cr
&=3(N-1)(2N+1).
&\eqname\cent\cr
}
$$

The number of super moduli is\refmark{\Ma}
$$
\eqalignno{
({\rm \#\  of\  super\  moduli})_{g,n}&=2(g-1)+n \cr
&\ +4(g-1)+2n \cr
&\ + \cdots \cr
&\ +2(N-1)(g-1)+(N-1)n\cr
&=N(N-1)(g-1)+N(N-1)n/2.
&\eqname\sm\cr
}
$$
Here we assume that all the punctures are NS states,
which is sufficient for later investigations.

\chapter{$N=1\  w_N\subset N=1\  w_{N+1} $}

In this section,
we show that there is a similar sequential embedding structure to
the $w_N$\ string in the $N=1$\ case.
For obtaining a special realization of
the $N=1$\ linear $w_{N+1}$\ algebra,
we must add the matter fields $(\beta_i(z),\gamma_i(z))$\ and
$(b_i(z),c_i(z)),i=1,\cdots,N-1$\ to the general critical
$N=1$\ superstring theory.
Here these additional fields have dimensions $(i+2,-i-1)$\ and
$(i+3/2,-i-1/2)$\ and
the superstring theory is represented by $N=1$\ super conformal generators
$(T(z),G(z))$\ with $c=15$.
A special realization of the $N=1$\ linear $w_{N+1}$\
algebra is given by
$$
\eqalignno{
w_i(z)&=\delta_{i,0}T+i\beta_i\cr
&+\sum^{N-i-1}_{j=1}
\Big(-(i+j+2)\beta_{i+j}\partial\gamma_j-
(j+1)\partial\beta_{i+j}\gamma_j
 -(i+j+{3\over2})b_{i+j}\partial c_j-
(j+{1\over2})\partial b_{i+j}c_j\Big),\cr
v_i(z)&=\delta_{i,0}G+ib_i\cr
&+\sum^{N-i-1}_{j=1}
\Big(2\beta_{i+j}c_j
-(i+j+{3\over2})b_{i+j}\partial\gamma_j-
(j+1)\partial b_{i+j}\gamma_j\Big).
&\eqname\Nlinw\cr
}
$$

The BRST operator is transformed into the sum of the BRST operators
of the $N=1\ w_N$\ string and the topological theory
by the similarity transformations:
$$
\eqalignno{
e^RQ_{BRST}^{N=1(N+1)}e^{-R}&=Q_{BRST}^{N=1(N)}+Q_{top},
&\eqname\tBRSTN\cr
}
$$
where
$$
\eqalignno{
Q_{top}&=\oint{dz\over2\pi i}\Big(
(N-1)c_{N-1}^{gh}\beta_{N-1}-(N-1)\gamma_{N-1}^{gh}b_{N-1}\Big).
&\eqname\Ntop\cr
}
$$
The generator of this similarity transformation is
$$
\eqalignno{
R={1\over N-1}\oint {dz\over 2\pi i}\sum^{N-2}_{i=0}\Big[
&c_i^{gh}\Big((N+1)b_{N-1}^{gh}\partial\gamma_{N-i-1}+
(N-i)\partial b_{N-1}^{gh}\gamma_{N-i-1}\cr
&\qquad
-(N+{1\over2})\beta_{N-1}^{gh}\partial c_{N-i-1}
-(N-i-{1\over2})\partial\beta_{N-1}^{gh}c_{N-i-1}\Big)\cr
-&\gamma_i^{gh}\Big(2b_{N-1}^{gh}c_{N-i-1}\cr
&\qquad+(N+{1\over2})\beta_{N-1}^{gh}\partial\gamma_{N-i-1}+
(N-i)\partial\beta_{N-1}^{gh}\gamma_{N-i-1}\Big)
\Big].
&\eqname\NRlin\cr
}
$$
Therefore the cohomology of the $N=1\ w_{N+1}$\ string coincides with
that of the $N=1\ w_N$\ string.

The coincidence of the amplitudes is easily shown by considering
the transformed theory.
We bosonize $(\beta_{N-1}^{gh}(z),\gamma_{N-1}^{gh}(z))$\
as
$$
\eqalignno{
\gamma_{N-1}^{gh}(z)&=e^{\phi^{gh}_{N-1}(z)}\eta^{gh}_{N-1}(z),\quad
\beta_{N-1}^{gh}(z)=\partial\xi^{gh}_{N-1}(z)e^{-\phi^{gh}_{N-1}(z)}.
&\eqname\bosogh\cr
}
$$
By means of these bosonized fields,
the physical vertex operators in the canonical picture can be factorized
to the physical vertex of the $N=1\ w_N$\ string
and of the topological theory.
$$
\eqalignno{
V^{(N+1)N=1}_{phys}(z)&=V^{(N)N=1}_{phys}(z)\Big(
\big(\prod^{N-1}_{n=0}\partial^nc_{N-1}(z)e^{-N\phi_{N-1}^{gh}(z)}\big)
\big(\prod^{N-1}_{n=0}\partial^nc_{N-1}^{gh}(z)e^{-N\phi_{N-1}(z)}\big)
\Big)\cr
&\equiv V^{(N)N=1}_{phys}(z)V^{top}(z).
&\eqname\Nphys\cr
}
$$
Thus the amplitudes of the $N=1\ w_{N+1}$\ string become the product
of those of the $N=1\ w_N$\ string and of the topological theory.

The picture changing operators
for calculating the topological amplitudes are
$$
\eqalignno{
Z_{N-1}(z)\equiv\{Q_{top},\xi_{N-1}(z)\}
&=(N-1)c^{gh}_{N-1}(z)e^{-\phi_{N-1}(z)},
&\eqname\pcN\cr
}
$$
for matter bosonic fields and
$$
\eqalignno{
Z_{N-1}^{gh}(z)\equiv\{Q_{top},\xi^{gh}_{N-1}(z)\}
&=(N-1)e^{\phi^{gh}_{N-1}(z)}b_{N-1}(z),
&\eqname\pcNgh\cr
}
$$
for ghost bosonic fields.
The inverse picture changing operators are easily found:
$$
\eqalignno{
Z_{N-1}^{-1}(z)&={1\over N-1}e^{\phi_{N-1}(z)}b_{N-1}^{gh}(z),\qquad
Z_{N-1}^{gh\ -1}(z)={1\over N-1}c_{N-1}(z)e^{-\phi_{N-1}^{gh}(z)}.
&\eqname\invpcN\cr
}
$$

The topological amplitudes are given by
$$
\eqalignno{
{\cal A}^{top}_{g,n}&=\VEV{
\prod^{(2N+1)(g-1)+Nn}_{i=1}Z^{-1}_{N-1}(z_i)
\prod^{2N(g-1)+Nn}_{i=1}Z^{gh}_{N-1}(z_i)
\prod^n_{k=1}V^{top}(z_k)
}\cr
&=1,
&\eqname\topamp\cr
}
$$
up to sign comes from interchanging the order of the fermionic operators.
The $Z^{-1}_{N-1}$\ ($Z^{gh}_{N-1}$) insertions come from
the spin $N+1$\ moduli (the spin $N+1/2$\ supermoduli) integrals.
Therefore the amplitudes of the $N=1\ w_{N+1}$\ string coincide
with those of the $N=1\ w_N$\ string.

\chapter{Another linearized $w_N$\ algebra}

So far we have shown that $(N=1)\ w_N$\ string theory can be obtained
as a special realization of the $(N=1)\ w_{N+1}$\ string.
However, this $w_N$\ string is not general but a special
as mentioned in \S 3 and \S 5.
Only the $w_2$\ (bosonic) string theory is included in the general form.
One may consider that this is insufficient and the general $w_N$\
string theory should be obtained
by a special realization of the $w_{N+1}$\
string. In this section, we explain that this is possible by
considering another linearization of the nonlinear $W_N$\ algebra.

The linearization of the $W_N$\ algebra is not unique
and there is another linearized $w_N$\ algebra
which was defined in Ref. [\hoso]:
$$
\eqalignno{
w_i(z)w_j(w)&\sim 0,\qquad {\rm for}\ i,j=1,\cdots,N-2,
&\eqname\abelianw\cr
}
$$
and the OPE related to $w_0(z)=T(z)$\ is the same with
that for the linear $w_N$\ algebra.
This can be obtained from the linear $w_N$\ algebra by a group contraction.
We call this linearized $w_N$\ algebra as abelian $w_N$\ algebra.

To obtain the BRST operator,
we have to replace the ghost generators in \linw\ to
$$
\eqalignno{
w_i^{gh}(z)&=-(i+2)b_i^{gh}\partial c_0^{gh}-\partial b_i^{gh}c_0^{gh},
\qquad {\rm for}\ i=1,\cdots,N-2.
&\eqname\abew\cr
}
$$

The general $w_N$\ string theory is represented by the generators
$\{\tilde w_i \},\ i=1,\cdots ,N-2$, with the critical central charge
$c(N)$ obtained in \cent.
A special realization of the abelian $w_{N+1}$\ algebra
is constructed by the matter system
composed of the above {\it general theory} with $w_N$\ symmetry
and the bosonic fields
$(\beta_{N-1}(z),\gamma_{N-1}(z))$, which have dimensions $(N+1, -N)$.
Generators of the $w_{N+1}$\ symmetry
$\{w_i\},\ i=0,\cdots ,N-1$\ is defined by
$$
\eqalignno{
w_i(z)&=\tilde w_i+
\delta_{i,0}\Big(-(N+1)\beta_{N-1}\partial\gamma_{N-1}-
N\partial\beta_{N-1}\gamma_{N-1}\Big),
\qquad{\rm for}\  i=0,\cdots ,N-2 \cr
w_{N-1}(z)&=(N-1)\beta_{N-1}.
&\eqname\asw\cr
}
$$
The BRST operator of this $w_{N+1}$\ string theory
is transformed into the same form
\BRSTone by the similarity transformation generated by
$$
\eqalignno{
R^{abelian}&=
{1\over N-1}\oint {dz\over 2\pi i}c_0^{gh}\Big((N+1)b_{N-1}^{gh}
\partial \gamma_{N-1}
+N\partial b_{N-1}^{gh}\gamma_{N-1}) \Big).
&\eqname\simone\cr
}
$$
The remaining proofs are the same with those of the linear $w_N$\ cases.

The $N=1$\ abelian $w_N$\ algebra
is similarly obtained by replacing the r.h.s. of
\Nw to zero in the case of $i\ne 0,j\ne 0$.
The ghost generators in BRST operator are now
$$
\eqalignno{
w^{gh}_i(z)&=-(i+2)b_i^{gh}\partial c_0^{gh}-\partial b_i^{gh}c_0^{gh}
-(i+{3\over2})\beta_i^{gh}\partial\gamma_0^{gh}
-{1\over2}\partial\beta_i^{gh}\gamma_0^{gh},\cr
v^{gh}_i(z)&=2b_i^{gh}\gamma_0^{gh}
-(i+{3\over2})\beta_i^{gh}\partial c_0^{gh}-\partial\beta_i^{gh}c_0^{gh},
&\eqname\ghwv\cr
}
$$
for $i=1,\cdots,N-2$.

A special realization of the $N=1$\ abelian $w_{N+1}$\ algebra is obtained
by adding the matter fields $(\beta_{N-1}(z),\gamma_{N-1}(z))$\ and
$(b_{N-1}(z),c_{N-1}(z))$\ to the system of the general critical
$N=1$\ abelian $w_N$\ string.
The dimensions of the additional fields are
$(N+1,-N)$\ and $(N+1/2,-N+1/2)$.
Then the special realization of the $N=1$\ abelian $w_{N+1}$\ algebra is
$$
\eqalignno{
w_i(z)&=\tilde w_i
+\delta_{i,0}\Big(-(N+1)\beta_{N-1}\partial\gamma_{N-1}-
\partial\beta_{N-1}\gamma_{N-1}\Big),\cr
v_i(z)&=\tilde v_i
+\delta_{i,0}\Big(2\beta_{N-1}\partial\gamma_{N-1}-
(N+{1\over2})b_{N-1}\partial\gamma_{N-1}-
N\partial b_{N-1}\gamma_{N-1}\Big),\cr
&&{\rm for}\ i=0,\cdots,N-2\cr
w_{N-1}(z)&=(N-1)\beta_{N-1},\cr
v_{N-1}(z)&=(N-1)b_{N-1},
&\eqname\Nabw\cr
}
$$
where $(\tilde w_i(z),\tilde v_i(z)),i=0,\cdots,N-2$\ are
generators of abelian $w_N$\
algebra. The transformed BRST operator has the same form with \tBRSTN\
if we replace the generator of the similarity transformation with
$$
\eqalignno{
R^{abelian}={1\over N-1}\oint {dz\over 2\pi i} \Big[
&c_0^{gh}\Big((N+1)b_{N-1}^{gh}\partial\gamma_{N-1}+
N\partial b_{N-1}^{gh}\gamma_{N-1}\cr
&\qquad-(N+{1\over2})\beta_{N-1}^{gh}\partial c_{N-1}-
(N-{1\over2})\partial\beta_{N-1}^{gh}c_{N-1}\Big)\cr
-&\gamma_0^{gh}\Big(2b_{N-1}^{gh}c_{N-1}
+(N+{1\over2})\beta_{N-1}^{gh}\partial\gamma_{N-1}+
N\partial\beta_{N-1}^{gh}\gamma_{N-1}\Big)
\Big].\cr
&&\eqname\NRab\cr
}
$$

We have shown that one can construct another $w_N$\ string theory based on
the abelian $w_N$\ algebra.
There is also embedding structure in the abelian $w_N$\ strings.
The general abelian $(N=1)\ w_N$\ string is obtained
as a special realization of
the abelian $(N=1)\ w_{N+1}$\ string.

\chapter{$w_N \subset N=1\  w_N$}

We can consider another embedding structure
since the $w_N$\ algebra can be embedded
also in the $N=1\ w_N$\ algebra.
In this section, we show that the $w_3$\ string theory
can be actually obtained as a special realization of
the $N=1\ w_3$\ string\rlap.\foot{
We should note that the linear and the abelian $w_N$\
algebras are identical for $w_3$.

}

A realization of the $N=1\ w_3$\ algebra is obtained by using
the $w_3$\ algebra generated by $(\tilde w_0(z),\tilde w_1(z))$\
and the fermionic fields $(b_0(z),c_0(z))$\ and
$(b_1(z),c_1(z))$\ with dimensions
$(3/2,-1/2)$\ and $(5/2,-3/2)$.
Generators of $N=1\ w_3$\ algebra is given by
$$
\eqalignno{
w_0(z)&=\tilde w_0 -{3\over2}b_0\partial c_o-{1\over2}\partial b_0c_0
-{5\over2}b_1\partial c_1-{3\over2}\partial b_1c_1
+{1\over2}\partial^2(c_0\partial c_0),\cr
w_1(z)&=\tilde w_1-2c_0\partial c_0\tilde w_1-{5\over2}b_1\partial c_0+
{1\over2}c_0\partial b_1,\cr
v_0(z)&=b_0+c_0\Big(\tilde w_0
-{5\over2}b_1\partial c_1-{3\over2}\partial b_1c_1-b_0\partial c_0\Big)
+7\partial^2c_0-{9\over8}c_0\partial c_0\partial^2c_0,\cr
v_1(z)&=b_1+2c_0\tilde w_1+{5\over2}c_0\partial c_0b_1.
&\eqname\Nwlin\cr
}
$$

The BRST operator can be transformed by the similarity transformation
generated by
$$
\eqalignno{
R=\oint {dz\over2\pi i}\Big[
&c_0^{gh}\Big(-{3\over2}\beta_0^{gh}\partial c_0
-{1\over2}\partial\beta_0^{gh}c_0
-{5\over2}\beta_1^{gh}\partial c_1-{3\over2}\partial\beta_1^{gh}c_1
-{1\over2}b_0^{gh}c_0\partial c_0\cr
&\qquad-{3\over2}\partial (b_1^{gh}c_0c_1)-b_1^{gh}c_0\partial c_1
-{35\over24}\beta_1^{gh}c_0\partial c_0\partial c_1
-{7\over8}\partial\beta_1^{gh}c_0\partial c_0c_1
-{5\over8}\beta_1c_0\partial^2c_0c_1\Big)\cr
&+c_1^{gh}\Big({5\over2}\beta_1^{gh}\partial c_0
+{1\over2}\partial\beta_1^{gh}c_0\Big)\cr
&+\gamma_0^{gh}\Big(-b_0^{gh}c_0
+{1\over4}\beta_0^{gh}c_0\partial c_0
+{5\over2}\partial (\beta_1^{gh}c_0c_1)-{1\over2}\partial\beta_1^{gh}c_0c_1
+{3\over4}b_1^{gh}c_0\partial c_0c_1\Big)\cr
&+\gamma_1^{gh}\Big(-2b_1^{gh}c_0
+{5\over4}\beta_1^{gh}c_0\partial c_0\Big)
+{5\over4}c_0\partial c_0b_1c_1\Big]
&\eqname\simsim\cr
}
$$
into the form:
$$
\eqalignno{
e^{-R}Q_{BRST}^{(3)N=1}e^R&=Q_{BRST}^{(3)}+Q_{top},
&\eqname\tBRSTNN
}
$$
where
$$
\eqalignno{
Q_{top}&=-\oint{dz\over2\pi i}\big(\gamma_0^{gh}b_0
+\gamma_1^{gh}b_1\big).
&\eqname\Ntoptwo\cr
}
$$
Thus the cohomology of $N=1\ w_3$\ string is equivalent to
the $w_3$\ string.
We note that the total stress tensor $w_0^{tot}(z)=w_0+w_0^{gh}$,
which has unusual total derivative term,
is transformed into the usual form\foot{
The total stress tensors in the previous cases,
which have no unusual term,
are invariant under the similarity transformations.
}:
$$
\eqalignno{
e^Rw_0^{tot}e^{-R}&=\tilde w_0
-{3\over2}b_0\partial c_0-{1\over2}\partial b_0c_0
-{5\over2}b_1\partial c_1-{3\over2}\partial b_1c_1\cr
&-2b_0^{gh}\partial c_0^{gh}-\partial b_0^{gh}c_0^{gh}
-3b_1^{gh}\partial c_1^{gh}-2\partial b_1^{gh}c_1^{gh}\cr
&-{3\over2}\beta_0^{gh}\partial\gamma_0^{gh}
-{1\over2}\partial\beta_0^{gh}\gamma_0^{gh}
-{5\over2}\beta_1^{gh}\partial\gamma_1^{gh}
-{3\over2}\partial\beta_1^{gh}\gamma_1^{gh}.
&\eqname\stress\cr
}
$$
The physical vertex operators in the transformed theory are
$$
\eqalignno{
V_{phy}^{(3)N=1}&=V_{phy}^{(3)}
\big(c_0(z)e^{-\phi_0^{gh}(z)}\big)\big(c_1(z)
\partial c_1(z)e^{-2\phi_1^{gh}(z)}\big).
&\eqname\phyverfin\cr
}
$$
The topological amplitudes can be calculated by using the picture changing
operators
$$
\eqalignno{
Z_i^{gh}(z)&\equiv\{Q_{top},\xi_i^{gh}(z)\}=e^{\phi_i^{gh}(z)}b_i(z),\cr
Z_i^{gh\ -1}(z)&=c_i(z)e^{-\phi_i^{gh}(z)},
&\eqname\pcNtwo\cr
}
$$
where $i=0,1$.
One can easily show the amplitudes of the $N=1\ w_3$\ string coincide
with those of the $w_3$\ string in a similar way to the previous cases.

Here we conjecture that the $w_N$\ string
is obtained as a special realization
of the $N=1\ w_N$\ string for general $w_N$.
If this is indeed the case,
the $N=1\ w_{\infty}$\ string is more universal than
the $w_{\infty}$\ string since it includes both sequences of $N=0$\
and $N=1\ w_N$\ string theories.
Such an investigation remains to the future study.

\chapter{Discussions}

We have constructed the $w_N$\ string theory
based on the linear $w_N$\ algebra. The linear $w_N$\ algebra
naturally tends to $w_{\infty}$, area preserving diffeomorphisms,
by taking the limit $N\rightarrow \infty$.
We have shown that there is sequential embedding structure:
the $w_N$\ string theory can be obtained
by a special realization of the $w_{N+1}$\ string.
This $w_N$\ string theory is, however, not general one but
it further gives $w_{N-1}$\ string theory\etc\ Only
the $w_2$\ (bosonic) string
is included as a whole.
The $w_{\infty}$\ string theory is a universal string theory
in this sequence.
It has been also shown that there
is a similar sequence for $N=1$\ string theories.
We conjecture that this sequence includes
the $N=0$\ string sequence
by explicit proof in the case of $w_3$\ string theory.
The $N=1\ w_{\infty}$\ string theory is therefore more universal.

We have also shown that the above discussions can be repeated
even if we replace the linear $w_N$\ algebra to the abelian $w_N$\ algebra
given by a contraction.
In this case, the general $w_N$\ string theory can be
obtained from the $w_{N+1}$\ string,
while $N\rightarrow\infty$\ limit does not
give the usual $w_{\infty}$\ algebra.
It gives a contraction of the area preserving
diffeomorphisms.

 From the results obtained in this paper,
we can expect that the $N=4\ w_{\infty}$\ string
theory is a candidate of the most universal string theory
since $N=4$\ supersymmetry is the maximum
in two dimensions and $w_{\infty}$\ is also the maximum extensions by using
the higher spin generators.
There is a fascinating possibility that all the string theories
can be obtained as a special realization
of the $N=4\ w_{\infty}$\ string theory.

It is interesting to study a geometrical picture of the symmetry breaking
considered in this paper.
A realization of the string theory is considered  to be
corresponding to a geometry of the target space.
We can ask what is the background geomety corresponding to the special
realization which is equivalent to the string based on a smaller symmetry.
We hope to discuss this issue elsewhere.

\endpage

\centerline{\bf Acknowledgements}

A part of the calculation in this paper
is performed by the package ope.math
for the operator product expansions developed by Akira Fujitsu,
whom the authors would like to thank for sending us his package.
One of the authors (H. K.) also acknowledge Katsumi Itoh and Yuji Igarashi
for valuable discussions.

\vskip 1cm

\refout

\bye